\documentstyle[pre,aps]{revtex}
\begin{document}
\draft
\title{Self-Similar Bootstrap of Divergent Series}

\author{V.I. Yukalov and S. Gluzman}
\address{International Centre of Condensed Matter Physics\\
University of Brasilia, CP 04513, Brasilia, DF 70919-970, \\
Brazil}
\maketitle
\begin{abstract}
A method is developed for calculating effective sums of divergent series.
This approach is a variant of the self-similar approximation theory. The
novelty here is in using an algebraic transformation with a power providing
the maximal stability of the self-similar renormalization procedure. The 
latter is to be repeated as many times as it is necessary in order to 
convert into closed self-similar expressions all sums from the series 
considered. This multiple and complete renormalization is called 
self-similar bootstrap. The method is illustrated by several examples 
from statistical physics. 

\end{abstract}


\section{Introduction}

The most powerfull analytic tool for solving realistic, and therefore hard,
problems in theoretical physics and applied mathematics is perturbation
theory. However, perturbation series are notoriously badly behaved : in the
majority of interesting cases they are divergent. A variety of mathematical
techniques have been invented to assign a finite value to the sum of a
divergent series. Such techniques are generally referred to as
renormalization or summation methods. These methods are not only useful to
theoretical and mathematical physicists but they are crucial because they
provide a way to recover physical information from perturbative
calculations. Probably the most common technique used to assign a meaningful
value to a divergent series is the Pade$^{\prime }$ summation [1]$.$ Using
the latter one converts a formal power series to a continued fraction.
Truncating this fraction at successive orders one obtains the rational
functions called Pade$^{\prime }$ approximants$.$ To reach a reasonable
accuracy of such approximants, one usually needs to have tens of terms in a
perturbative series.

Recently, a method has been suggested [2-4] permitting to ascribe meaningful
values to the limits of divergent sequences by exploiting just a few terms
of perturbative series. This approach, called self-similar approximation
theory [2-4]$,$ is based on the following ideas.

First, one has to incorporate into the considered sequence additional
functions whose role is to renormalize the sequence making it convergent.
These functions, because of their role, are named governing or control
functions. The latter are to be defined from fixed-point conditions [2-4]$.$
There are several ways of introducing such control functions. One and a
natural way is to include them into initial approximation [5,6]$.$
Fixed-point equations may be written in the form of the minimal-difference
condition [5-7]$,$ or the minimal-sensitivity condition [8-14]$.$ A
condition close to the latter type has also been used in the potential
envelope method [15,16]$.$ In two simple cases, of a zero-dimensional and
one-dimensional anharmonic oscillator, the control functions were found
analytically for an arbitrary perturbation order [17-20] by requiring the
convergence of renormalized perturbation theory. Another way of introducing
control functions is by using a scaled basis [21-24]$,$ in which the scaling
parameters, playing the role of such functions, are given by analytical
expressions with coefficients adjusted empirically from the convergence of a
numerical iterative procedure. One can also define scaling parameters
without using their analytical representation, directly from numerical
search providing the convergence of an iterative method [25-29]$.$ In the
present paper we suggest a new way of introducing control functions,
different from all variants mentioned.

Another step in the self-similar approximation theory is to construct an
approximation cascade whose trajectory is bijective to the approximation
sequence considered [30-32]$.$ In this paper we construct such a cascade not
for the sequence itself but for its transform. Of course, the idea of
considering a transformed series instead of the initial one is not new. This
is, e.g., the basis of the known Borel summation. Another example is the use
of the Chebyshev transforms instead of straightforward power-series
representation [33]$.$ What is new in our approach is the use of a power-low
algebraic transform with powers playing the role of control functions.

Then the constructed approximation cascade is embedded into an approximation
flow. Integrating the evolution equation of the flow, we obtain a
self-similar approximation [2-4, 30-32]$.$

In this paper we define the control functions from the principle of maximal
stability of the approximation cascade trajectories. This is done by
minimizing the absolute values of mapping multipliers, which is equivalent
to a quasifixed point condition [30-32]$,$ since multipliers tend to zero
when approaching a stable fixed point.

The plan of the paper is as follows. In Sec.2{\rm \ }the algebraic
transforms are introduced and the main steps of the self-similar
approximation theory are sketched, not going into mathematical details which
can be found in Refs.2-4, 30-32, 34. In Sec.3 the procedure of self-similar
bootstrap is defined, consisting of multiple self-similar renormalization of
all sums entering into given series. In Sec.4 a particular case of the
self-similar bootstrap is considered leading to a nice representation in the
form of multiple exponentials. The following Sections illustrate the
approach by various examples emphasizing the generality of the method that
can be applied to problems of quite different nature.

\section{Algebraic Transforms}

Assume that we are interested in finding a function $f(x)$ of a real
variable $x\in (-\infty ,+\infty ).$ Without loss of generality, the
function $f(x)$ may be considered to be real, since the case of a complex
function can be always reduced to that of two real functions. Let
perturbation theory give for the function $f(x)\ $approximations $p_k(x)$
with $k=0,1,2,..$. enumerating the approximation number. The standard form
of $p_k(x)$ is a series in powers, not necessarily integer, of $x$. The
series can even include logarithms, since the latter, can always be
presented, using the replica trick, as an expression containing a noninteger
power.

The algebraic transform is defined as 
\begin{equation}
\label{1}P_k(x,s)=x^sp_k(x), 
\end{equation}
with $s$ real. This transform changes the powers of the series $p_k(x)$ 
changing by this the convergence properties of the latter.
Effectively, the approximation order increases from $k\ $to $k+s$ as a
result of Eq.(1). The transform inverse to (1) is 
\begin{equation}
\label{2}p_k(x)=x^{-s}P_k(x,s). 
\end{equation}

To construct an approximation cascade, we proceed as follows. Define the
expansion function $x=x(f,s)$ by the equation 
\begin{equation}
\label{3}P_0(x,s)=f, 
\end{equation}
where $P_0$ is the first available expression from Eq.(1). Substituting $%
x(f,s)$ back into (1), we get 
\begin{equation}
\label{4}y_k(f,s)\equiv P_k(x(f,s),s). 
\end{equation}
The left -hand side of (4) represents a point of the approximation-cascade
trajectory corresponding to approximation (1). The transformation inverse to
(4) reads 
\begin{equation}
\label{5}P_k(x,s)=y_k(P_0(x,s),s). 
\end{equation}

The function (4) realizes the endomorphism
$$
y_k(f,s):{\bf R\times R_{+}\rightarrow R\ .} 
$$
Consider the family $\{y_k:k\in {\bf Z_{+}}\}$ as a dynamical system in
discrete time. Since the trajectory of this dynamical system, according to
(4) and (5), is bijective to the approximation sequence $\{P_k\},$ this
system was called [30-32] the approximation cascade. In order to simplify
the consideration, let us pass from discrete time to continuous one. To this
end, embed the approximation cascade into an approximation flow:%
$$
\{y_k:k\in {\bf Z_{+}}\}\subset \{y(t,...):t\in {\bf R_{+}}\}, 
$$
which means that the trajectory $\{y(t,f,s)\}$ of the flow has to pass
through all points $\{y_k(f,s)\}$ of the cascade trajectory.

The evolution equation 
\begin{equation}
\label{6}\frac \partial {\partial t}y(t,f,s)=v(y(t,f,s),s) 
\end{equation}
for the approximation flow, where $v(f,s)$ is the velocity field, can be
integrated for an arbitrary time interval, say, from $t=k-1\ $to $t=k^{*},$
which gives 
\begin{equation}
\label{7}\int_{y_{k-1}}^{y_k^{*}}\frac{df}{v(f,s)}=k^{*}-k+1; 
\end{equation}
here 
$$
y_k=y(k,f,s),\ \ y_k^{*}=y(k^{*},f,s). 
$$
The upper limit in (7) corresponds, according to (5), to an approximation 
\begin{equation}
\label{8}P_k^{*}(x,s)=y(k^{*},P_0(x,s),s)\ . 
\end{equation}
The moment $t=k^{*}$ is chosen so that to reach the approximation (8) by the
minimal number of steps. That is, we require that the right-hand side of (7)
be minimal, 
\begin{equation}
\label{9}t_k^{*}\equiv \min (k^{*}-k+1)\ . 
\end{equation}
Under condition (9), expression (8) is called the self-similar approximation.

To find (8) explicitly, we need to concretize in (7) the velocity field $%
v(f,s).$ This can be done by the Euler discretization of (6) yielding the
finite difference 
\begin{equation}
\label{10}v_k(f,s)=y_k(f,s)-y_{k-1}(f,s). 
\end{equation}
Thus, using (5), the evolution integral (7) can be written as 
\begin{equation}
\label{11}\int_{P_{k-1}}^{P_k^{*}}\frac{df}{v_k(f,s)}=t_k^{*}\ , 
\end{equation}
where 
$$
P_k=P_k(x,s),\quad P_k^{*}=P_k^{*}(x,s)\ . 
$$

When no additional restrictions are imposed, the minimal number of steps for
reaching a quasifixed point is, evidently, one, 
\begin{equation}
\label{12}abs\min \ t_k^{*}=1. 
\end{equation}
Additional restrictions can be of different types. For example, if the value
of the sought function at some point $x_0$ is known, we may require that the
found approximation would coincide at this point with the given exact value.
Looking for an approximation in the class of functions with a prescribed
symmetry is another way of imposing restrictions. In some cases the
asymptotic behavior of the sought function at $x\rightarrow 0$ and $%
x\rightarrow \infty $ may be available. Then requiring the correct
asymptotic properties also plays the role of such additional constraints. In
what follows, we shall concretize these variants by illustrating them with
explicit examples.

In this way, the sole quantity that is not yet defined is the parameter $s$
of the transformation (1). Recall that our aim is to find an approximate
fixed point of the cascade trajectory, a quasifixed point, which, by
construction, represents the sought function. Therefore, the power $s$ of
the transform in (1) is to be chosen so that to force the trajectory of the
approximation dynamical system to approach an attracting fixed point. Recall
that $s$ here is nothing but a kind of the control function, so, it is to be
defined by a fixed-point condition. As is discussed in the Introduction,
there are several forms of equations defining fixed points. Here we opt for
a condition following from the analysis of fixed-point stability properties.

Considering the mapping given by the approximation cascade, we may introduce
the mapping multipliers 
\begin{equation}
\label{13}\mu _k(f,s)=\frac \partial {\partial f}\ y_k(f,s)\ . 
\end{equation}
This is related to the local Lyapunov exponent $\lambda _k$ through the
formula $\lambda _k=\frac 1k\ln \left| \mu _k\right| .$ Consequently, $\mu
_k\sim e^{\lambda _kk}.$ If at increasing time, here at $k\rightarrow \infty 
$$,\ $the trajectory approaches an attracting fixed point, then $\lambda
_k\rightarrow \lambda <0.$ This implies that the multiplier $\mu
_k\rightarrow 0,$ as $k\rightarrow \infty .$ Another quantity related to the
multiplier (13) is the predictability time [35] which can be defined as $%
\tau _k\approx \left| \lambda _k\right| ^{-1}$, or $\tau _k\approx \left|
k/\ln \left| \mu _k\right| \right| .$ This is the characteristic time during
which the motion along the cascade trajectory effectively approaches a fixed
point. When the latter is attractive, that is, when the limit of the local
Lyapunov exponent $\lambda _k,\ $as $k\rightarrow \infty ,\ $tends to a
negative value $\lambda <0,\ $then $\mu _k$ tends to zero, and at the same
time, larger absolute values $\left| \lambda _k\right| $ lead to smaller
characteristic times $\tau _k.$

These properties show that closer we are to a fixed point, smaller is the
absolute value of the multiplier (13). Hence, we may define the control
function $s$ as that providing the minimum of the multiplier. Instead of the
multiplier (13), as a function of the variable $f,$ it may be more
convenient to pass to its image 
\begin{equation}
\label{14}m_k(x,s)=\mu _k(F_0(x,s),s) 
\end{equation}
being a function of the variable $x.$ Then the control function $s=s_k(x)$
is defined by the equation 
\begin{equation}
\label{15}\left| m_k(x,s_k(x))\right| =\min _s\left| m_k(x,s)\right| . 
\end{equation}
Because the minimization of the multiplier makes the trajectory more stable,
we can call Eq.(15) {\it \ the principle of maximal stability. }And the so
defined control function $s_k(x)$ can be termed the {\it stabilizing} {\it %
control function}, or, for brevity, the {\it stabilizer}. Notice, for
comparison, that another definition of the fixed point would be to require
the velocity (10) to be zero, which is exactly the minimal-difference
condition [5,6]. After the stabilizer $s_k(x)$ is found from (15), we
substitute it into (8) and, using the inverse transformation (2), we obtain
the{\it \ self-similar approximation 
\begin{equation}
\label{16}f_k^{*}(x)=x^{-s_k(x)}P_k^{*}(x,s_k(x)) 
\end{equation}
}for the sought function{\it .}

At the end of this section, let us note that the choice of control functions
from fixed-point equations is rigorously justified when the fixed point is
stable, that is if $\left| \mu _k\right| <1.$ When at some point $k$ the
trajectory becomes unstable, we have to stop the calculational procedure at
the last stable point. Another possibility is to restructure the considered
perturbation series. For instance, the instability of the procedure often
happens when we are trying to construct self-similar approximation for a
divergent function. Assume that we are dealing with such a function $f(x)$
which diverges as $x\rightarrow \infty .$ Then, the self-similar
renormalization procedure may become unstable at large $x$. To avoid the
instability, we can\ either consider the function $f^{-1}(x)$ or can rewrite
the series in powers of $1/x$, treating the latter as a new expansion
parameter. Usually, after this reexpansion the stability is restored. In the
following, we shall illustrate this possibility by an example and will
suggest a simple trick, giving the answer without the reexpansion, although
being equivalent to the latter.

\section{Self-Similar Bootstrap}

The procedure of calculating the self-similar approximations (16), starting
from a perturbative series $p_k(x)$ is now completely defined. The
renormalized quantity $f_k^{*}(x)$ must be, by construction involving the
stability properties, a much better approximation to the sought function $%
f(x)$ than the initial perturbative series $p_k(x).$ To improve the
accuracy, we may repeat the self-similar renormalization applying it to
other series that are left in (16).

For illustrating this {\it multiple self-similar renormalization, }consider
explicitly a perturbative series 
\begin{equation}
\label{17}p_k(x)=\sum_{n=0}^ka_nx^n,\quad a_0\neq 0, 
\end{equation}
containing integer powers of $x,$ although, as is mentioned above, the
procedure works for arbitrary noninteger powers. Following Sec.2, write the
algebraic transform 
\begin{equation}
\label{18}P_k(x,s)=\sum_{n=0}^ka_nx^{n+s} 
\end{equation}
of (17). As is seen, the transform (18) corresponds to an effectively higher
perturbation order, $k+s$, as compared to the initial series (17), of order $%
k$. Eq.(3) for the expansion function $x(f,s)$ now reads 
\begin{equation}
\label{19}P_0(x,s)=a_0x^s=f\ , 
\end{equation}
from where 
\begin{equation}
\label{20}x(f,s)=(\frac f{a_0})^{1/s}. 
\end{equation}
The cascade-trajectory points in (4) become 
\begin{equation}
\label{21}y_k(f,s)=\sum_{n=0}^ka_n(\frac f{a_0})^{n/s+1}. 
\end{equation}
The velocity field (10) writes 
\begin{equation}
\label{22}v_k(f,s)=a_k(\frac f{a_0})^{1+k/s}. 
\end{equation}
Calculating the evolution integral (11), with condition (12), we get the
approximation (8) in the form 
\begin{equation}
\label{23}P_k^{*}(x,s)=P_{k-1}(x,s)[1-\frac{ka_k}{sa_0^{1+k/s}}%
P_{k-1}^{k/s}(x,s)]^{-s/k}. 
\end{equation}
The stabilizer $s_k(x)$ is to be found from the minimization of the
multiplier 
\begin{equation}
\label{24}m_k(x,s)=\sum_{n=0}^k\frac{a_n}{a_0}(1+\frac ns)\ x^n 
\end{equation}
from (14). Then we obtain the self similar approximation (16), 
\begin{equation}
\label{25}f_k^{*}(x)=p_{k-1}(x)[1-\frac{ka_k}{sa_0^{1+k/s}}%
x^kp_{k-1}^{k/s}(x)]^{-s/k}\ , 
\end{equation}
where $s=s_k(x).$

In this way, the self-similar renormalization led us from the initial
perturbative series (17) to the renormalized approximation (25). The latter
contains a perturbative series $p_{k-1}(x)$ of the lower order than the
initial $p_k(x).$ This can be written as the relation 
\begin{equation}
\label{26}f_k^{*}(x)=F_k(x,p_{k-1}(x)) 
\end{equation}
showing that (25) depends on $x$ and $p_{k-1}(x).$ We may repeat the
procedure renormalizing $p_{k-1}(x)$ and getting $f_{k-1}^{*}(x).$ With such
a double renormalization, we come from (25) to 
\begin{equation}
\label{27}f_k^{**}(x)=F_k(x,\ f_{k-1}^{*}(x))=F_k(x,F_{k-1}(x,p_{k-2}(x))). 
\end{equation}
Here the doubly renormalized $f_k^{**}(x)\ $is expressed through $%
p_{k-2}(x). $ Repeating the self-similar renormalization $k$ times, we
obtain the $k$-fold self-similar approximation 
\begin{equation}
\label{28}f_k^{*...*}(x)=F_k(x,F_{k-1}(x,...(F_1(x,a_0))...)), 
\end{equation}
where we took into account that $p_0(x)=a_0.$ It may happen that (28)
contains other power series. Then we may renormalize them as well. When all
power series are renormalized, so that none of them is left unrenormalized,
we have 
\begin{equation}
\label{29}f_k^{*...*}(x)\rightarrow \widetilde{f_k}(x). 
\end{equation}
This complete procedure of the self-similar renormalization of all power
series, resulting in an expression $\widetilde{f_k}(x)$ containing none of
them, will be called the {\it self-similar bootstrap.}

\section{Multiple Exponentials}

There are particular cases of the multiple self-similar renormalization
yielding to a nice exponential representation of the resulting formulas.
Here we consider one such a sufficient condition when all coefficients at
perturbative powers are positive, $a_n>0.$ Then the minimum of the
multiplier (24) realizes at $s\rightarrow \infty .$ Taking this limit in
(25) gives 
\begin{equation}
\label{30}f_k^{*}(x)=p_{k-1}(x)\exp (\frac{a_k}{a_0}x^k). 
\end{equation}
Repeating the renormalization in line with (26) and (27), we get 
\begin{equation}
\label{31}f_k^{**}(x)=p_{k-2}(x)\exp \{\frac
1{a_0}(a_{k-1}x^{k-1}+a_kx^k)\}. 
\end{equation}
Continuing so on, we obtain the $k$-fold approximation (28) in the form 
\begin{equation}
\label{32}f_k^{*...*}(x)=a_0\exp \{\frac 1{a_0}(a_1x+a_2x^2+...+a_kx^k)\}. 
\end{equation}
As we see, the $k$-times repeated self-similar renormalization does not
deliver us from the power series. Really, the $k$-star approximation (32) is
expressed through a part of the initial perturbation series (17), namely,
through 
$$
p_k(x)-a_0=\sum_{n=1}^ka_nx^n. 
$$
With the notation 
\begin{equation}
\label{33}p_k^{^{\prime }}(x)\equiv \sum_{n=0}^ka_n^{^{\prime }}x^n, 
\end{equation}
in which $a_n^{^{\prime }}\equiv a_{n+1,}$ $n=0,1;2,...k,$ we may rewrite
(32) as 
\begin{equation}
\label{34}f_k^{*...*}(x)=a_0\exp \{\frac x{a_0}p_{k-1}^{^{\prime }}(x)\}. 
\end{equation}
The power series $p_{k-1}^{^{\prime }}(x)$ can be renormalized in our
standard way giving the corresponding self-similar approximation 
\begin{equation}
\label{35}f_{k-1}^{^{\prime }}(x)=a_0^{^{\prime }}\exp \{\frac
x{a_0^{^{\prime }}}p_{k-2}^{^{\prime \prime }}(x)\}, 
\end{equation}
in which 
\begin{equation}
\label{36}p_k^{^{\prime \prime }}(x)\equiv \sum_{n=0}^ka_n^{^{\prime \prime
}}x^n,\quad a_n^{^{\prime \prime }}\equiv a_{n+2}\ . 
\end{equation}
With this renormalization in mind, we transform (34) into 
\begin{equation}
\label{37}f_k^{*...*}(x)=a_0\exp \{\frac x{a_0}f_{k-1}^{^{\prime }}(x)\}. 
\end{equation}
Combining (35) and (37), we have 
\begin{equation}
\label{38}f_k^{*...*}(x)=a_0\exp \{\frac x{a_0}a_1\exp \{\frac
x{a_1}p_{k-2}^{^{\prime \prime }}(x)\}\}. 
\end{equation}
Converting $k$ times all power series in the exponentials, with the use of
the notation 
\begin{equation}
\label{39}b_0=a_{0\ },\quad b_k=\frac{a_k}{a_{k-1}}\ ,\ \ k=1,2,..., 
\end{equation}
we obtain the {\it bootstrap self-similar approximation 
\begin{equation}
\label{40}\widetilde{f_k}(x)=b_0\exp (b_1x\exp (b_2x\exp (...b_{k-1}x\exp
(b_kx)))...), 
\end{equation}
}as is discussed in (29).

In the case of small $x\rightarrow 0$ the expression (40) yields 
\begin{equation}
\label{41}\widetilde{f_k}(x)\simeq c_0+c_1x+c_2x^2+c_3x^3 
\end{equation}
with the coefficients 
$$
c_0=b_0,\quad c_1=b_0b_1,\quad c_2=b_0b_1(b_2+\frac 12b_1),\quad
c_3=b_0b_1(b_2b_3+\frac 12b_2^2+b_1b_2+\frac 16b_1^2). 
$$
Substituting here (39), we have 
$$
c_0=a_0,\quad c_1=a_1,\quad c_2=a_2+\frac{a_1^2}{2a_0},\quad c_3=a_3+\frac{%
a_2^2}{2a_1}+\frac{a_1a_2}{a_0}+\frac{a_1^3}{6a_0^2}\ \ . 
$$
This shows that the asymptotic, as $x\rightarrow 0,\ $behavior of (41) and
(17) coincides up to the linear terms. For the higher-order terms, $a_n\neq
b_n$ for $n\geq 2.$ Such a renormalization of the higher order expansion
coefficients is typical of the self-similar approximation theory [2-4]. This
renormalization allows to extend the region of applicability of self-similar
approximations, with respect to a variable $x,\ $as compared to the initial
perturbative series.

It is worth mentioning that the multiple, or continued, exponentials of type
(40) have been studied in mathematical literature for already more than two
centuries, since Euler; and a number of references can be found in Refs.
[36,37]. We have derived the form (40) following the multiple self-similar
renormalization for the power series (17) with positive coefficients. Of
course, for other sets of coefficients in (17) the final bootstrap
approximation will not necessarily take the form of a multiple exponential,
as in (40), but will be a mixture of exponentials and fractions, each
expression being conditioned by the principle of maximal stability (15).

Another way of getting a multiple exponential could be as follows. Consider
a sequence $\{\varphi _k(x)\}$ of the terms%
$$
\varphi _1(x)=b_0\exp (b_1x), 
$$
$$
\varphi _2(x)=b_0\exp (b_1x\exp (b_2x)), 
$$
and so on, where the coefficients $b_k$ are given by (39). At the $k$-step
of the sequence $\{\varphi _k(x)\}$ we come to (40). However, this way is
justified if the approximation cascade corresponding to the sequence $%
\{\varphi _k(x)\}$ possesses a stable trajectory leading to a quasifixed
point representing (40). This approximation cascade can be constructed in
the standard way [32,34]. From the equation $\varphi _1(x)=f$ we find $%
x=x(f) $ which is 
$$
x(f)=\frac{a_0}{a_1}\ln (\frac f{a_0})\ . 
$$
Then we define the approximation cascade as is described in Sec.2. The
corresponding trajectory $\{z_k(f)\},$ consisting of the terms 
$$
z_k(f)\equiv \varphi _k(x(f)) 
$$
is bijective to the approximation sequence $\{\varphi _k(x)\}.$

The stability of the trajectory $\{z_k(f)\}$ is checked by calculating the
multipliers 
\begin{equation}
\label{42}M_k(x)\equiv [\frac{\partial z_k(f)}{\partial f}]_{f=\varphi
_1(x)} 
\end{equation}
and analyzing them with respect to the stability condition $\left|
M_k(x)\right| <1$.

At the end of this section let us remark that we have used here the word
''bootstrap'' in the generally accepted meaning, as a kind of a completely
self-consistent procedure permitting to construct an explicit solution to a
complicated problem. This term in the close meaning was used, for example,
in constructing a self-consistent distribution over particle masses in
high-energy physics [38], and also in constructing the $S$-matrices for
two-dimensional conformal field theories [39].

\section{Zero-Dimensional Analogs of Field Theories $\ $}

\subsection{Non-Degenerate Vacuum}

Consider the thermodynamic potential of a zero-dimensional anharmonic model
(see e.g [32]) represented by the integral 
\begin{equation}
\label{43}J(g)=\frac 1{\sqrt{\pi }}\int_{-\infty }^\infty \exp
(-x^2-gx^4)dx, 
\end{equation}
with the integrand possessing a single ''vacuum'' state, located at the
point $x=0\ $.\ The expansion of this integral in powers of the coupling
parameter $g$, around the vacuum\ state leads to divergent series, 
\begin{equation}
\label{44}J(g)\sim a+bg+cg^2+dg^3+hg^4+..., 
\end{equation}
$$
a=1,\ b=-\frac 34,\ c=\frac{105}{32},\ d=-\frac{3465}{128},\ h=\frac{675675}{%
2048}\ . 
$$

We apply here the self-similar bootstrap renormalization, guided by desire
to perform as many renormalization steps, as possible. Write down the
following set of approximations to the quantity $J(g)$, analogous to the
general form (17): 
$$
J_0(g)=a, 
$$
$$
J_1(g)=a+bg, 
$$
\begin{equation}
\label{45}J_2(g)=a+bg+cg^2
\end{equation}
$$
J_3(g)=a+bg+cg^2+dg^3, 
$$
$$
J_4(g)=a+bg+cg^2+dg^3+hg^4, 
$$
together with the following local multipliers, that can be found from the
general representation (24):%
$$
m_1(g,s)=1+\frac ba\frac{1+s}sg\ , 
$$
$$
m_2(g,s)=m_1(g,s)+\frac ca\frac{2+s}sg^2, 
$$
\begin{equation}
\label{46}m_3(g,s)=m_2(g,s)+\frac da\frac{3+s}sg^3,
\end{equation}
$$
m_4(g,s)=m_3(g,s)+\frac ha\frac{4+s}sg^4. 
$$
The analysis of (46) shows, that in the last three cases the most stable
trajectories are realized at $s\rightarrow \infty ,\ $and that in the first
case, $s\rightarrow \infty $ also corresponds to a stable trajectory.
Therefore the starting four steps of the self-similar bootstrap
renormalization can be safely performed in the exponential form, leading to
the intermediate renormalized quantity 
\begin{equation}
\label{47}J_4^{****}(g)=a\exp \{\frac 1a(bg+cg^2+dg^3+hg^4)\}.
\end{equation}
Write down a set of approximations to the quantity $J^{\prime}(g)=bg+cg^2+dg^3+hg^4,\ $appearing in the exponential of this expression: 
$$
J_1^{^{\prime }}(g)=bg, 
$$
$$
J_2^{^{\prime }}(g)=bg+cg^2, 
$$
$$
J_3^{^{\prime }}(g)=bg+cg^2+dg^3, 
$$
$$
J_4^{^{\prime }}(g)=bg+cg^2+dg^3+hg^4, 
$$
with the following local multipliers:%
$$
m_2^{\prime }(g,s)=1+\frac cb\frac{2+s}{1+s}g, 
$$
\begin{equation}
\label{48}m_3^{\prime }(g,s)=m_2^{\prime }(g,s)+\frac db\frac{3+s}{1+s}g^2,
\end{equation}
$$
m_4^{\prime }(g,s)=m_3^{\prime }(g,s)+\frac hb\frac{4+s}{1+s}g^3. 
$$
The analysis of (48) leads us to the conclusion that the exponential
renormalization is optimal at every step and, following to the standard
prescriptions of Sec. 4, we transform (47) into 
\begin{equation}
\label{49}J_4^{****}(g)=a\exp \{\frac bag\exp [\frac 1b(cg+dg^2+hg^3)]\}\ .
\end{equation}
Our routine procedure requires to renormalize now the quantity $J^{^{\prime
\prime }}(g)=cg+dg^2+hg^3,\ $using the following approximations:%
$$
J_1^{^{\prime \prime }}(g)=cg, 
$$
$$
J_2^{^{\prime \prime }}(g)=cg+dg^2, 
$$
$$
J_3^{^{\prime \prime }}(g)=cg+dg^2+hg^3, 
$$
and analyzing the following multipliers:%
$$
m_2^{\prime \prime }(g,s)=1+\frac dc\frac{2+s}{1+s}g, 
$$
$$
m_3^{\prime \prime }(g,s)=m_2^{\prime \prime }(g,s)+\frac hc\frac{3+s}{1+s}%
g^2. 
$$
We conclude that the most stable trajectory corresponds, in both cases, to
the exponential summation, leading to the intermediate formula 
\begin{equation}
\label{50}J_4^{****}(g)=a\exp \{\frac bag\exp [\frac cb\exp [\frac
1c(dg+hg^2)]]\}.
\end{equation}
The last step of the procedure, applied to the quantity $J^{^{\prime \prime
\prime }}(g)=dg+hg^2$, with the approximations set:%
$$
J_1^{^{\prime \prime \prime }}(g)=dg, 
$$
$$
J_2^{^{\prime \prime \prime }}\text{(}g)=dg+hg^2, 
$$
and with the multiplier 
$$
m_2^{\prime \prime \prime }(g,s)=1+\frac hd\frac{2+s}{1+s}g, 
$$
again should be performed with $s\rightarrow \infty $, and the bootstrap
program is completed: 
\begin{equation}
\label{51}\widetilde{J_4}(g)=a\exp \{\frac bag\exp [\frac cb\exp [\frac
dcg[\exp (\frac hdg)]]]\}.
\end{equation}
Similar expressions follow when less terms from the initial expansion are
taken into account:%
$$
\widetilde{J_3}(g)=a\exp \{\frac bag\exp [\frac cb\exp (\frac dcg)]\}, 
$$
$$
\widetilde{J_2}(g)=a\exp \{\frac bag\exp [\frac cbg]\}, 
$$
$$
\widetilde{J_1}(g)=a\exp \{\frac bag\}. 
$$

We had pointed out already, that the last expression corresponds to the
stable, but not to an optimal, i.e. the most stable, trajectory. Analyzing $%
m_1(g,s)$, we obtain the optimally renormalized expression: 
\begin{equation}
\label{52}\widetilde{J_{1o}}(g)=a[1-\frac b{a\times s(g)}g]^{-s(g)},\quad
s(g)=\frac{-bg}{a+bg}. 
\end{equation}
At the point $g=1$,$\ $the following numbers are generated by the sequence $
\widetilde{J_i},\ i=2,3$...:%
$$
\widetilde{J_1}(1)=0.472\ (\widetilde{J_{1o}}{}=0.512),\quad \widetilde{J_2}%
(1)=0.991,\ \ \widetilde{J_3}(1)=0.473,\ \ \widetilde{J_4}(1)=0.991. 
$$
We observe the two subsequences, with odd and even numbers, with values
practically unchanged within each subsequence, probably embracing the
correct result from below and from above, respectively. We can suspect that
the corresponding sequence $\widetilde{J_i}\ $\ possesses the two competing
unstable quasifixed points, i.e behaves chaotically and, in such situation,
it is appropriate to use the self-similar approximation smoothed by the
Cesaro averaging procedure[32], i.e. in our case, simply to take the average
over the two neighboring members of each subsequence.

This conjecture, is supported by the analysis of the corresponding sequence
of the multipliers for the sequence $\widetilde{J_i}\ $, as discussed in
Sec.4. From the initial approximation 
$$
\widetilde{J_1}(g)=f, 
$$
one can find the expansion function 
$$
g=\frac ab\ln (\frac f{\sqrt{\pi }a}), 
$$
and, after the routine transformations, the following expression for the
multipliers (42) can be obtained:%
$$
M_1(g)\equiv 1, 
$$
$$
M_2(g)=\Phi _2(g)\Psi _2(g) 
$$
$$
M_3(g)=\Phi _3(g)\Psi _3(g), 
$$
$$
M_4(g)=\Phi _4(g)\Psi _4(g), 
$$
where 
$$
\Phi _2(g)=\frac{\widetilde{J_2}(g)}a\exp (\frac cbg), 
$$
$$
\Phi _3(g)=\frac{\widetilde{J_3}(g)}a\exp (\frac cbg\exp (\frac dcg)), 
$$
$$
\Phi _4(g)=\frac{\widetilde{J_4}(g)}a\exp (\frac cbg\exp (\frac dcg\exp
(\frac hdg))), 
$$
and%
$$
\Psi _2(g)=b+cg, 
$$
$$
\Psi _3(g)=b+cg\exp (\frac dcg)+dg^2\exp (\frac dcg), 
$$
$$
\begin{array}{c}
\Psi _4(g)=b+cg\exp (\frac dcg\exp (\frac hdg))+dg^2\exp (\frac hdg)\exp
(\frac dcg\exp (\frac hdg))+ \\ 
+hg^3\exp (\frac hdg)\exp (\frac dcg\exp (\frac hdg)). 
\end{array}
$$

The following values are obtained at $g=1$:%
$$
M_1=1,\quad M_2=-0.089,\quad M_3=1.008,\quad M_4=-0.089, 
$$
supporting our initial guess, that the approximation cascade behaves
chaotically. After the Cesaro averaging, the sought value at $g=1$ equals
to, say, $\frac{\widetilde{J_3}(1)+\widetilde{J_4}(1)}2=0.731$, deviating
from the exact value $0.772$ with the percentage error $-5.228\%$, an
acceptable accuracy if to remember that the initial expansion (44) gives the
percentage error $\sim $$10^4\%$. With the optimized $\widetilde{J_{1o}}$,${%
\ }$we obtain even better estimate, $0.752$, for the sought value, with the
percentage error equal to $-2.668\%$.

\subsection{Double-Degenerate Vacuum}

Consider the integral 
\begin{equation}
\label{53}I(g)=\int_{-\infty }^\infty \exp (x^2-gx^4)dx, 
\end{equation}
representing zero-dimensional field-theory with the integrand possessing the
two maxima, located at the points%
$$
\overline{x}=\pm \frac 1{\sqrt{2g}}, 
$$
where $g$ plays the role of a coupling constant. We intend to estimate this
integral in the region of intermediate couplings $g\sim 1.$ It was pointed
out in [40], that any conventional expansion, in powers of $g,\ $or $%
g^{-1},\ $is not sufficient, since it does not take into account the
existence of those degenerate maxima, corresponding to the double-degenerate
''vacuum''.

Within the framework of $D-$dimensional field theories, the existence of
degenerate vacuum is taken into account, e.g., by means of the zero-energy
instanton-anti-instanton solutions, and all further corrections to
observables come from the excitations above the instanton-anti-instanton
background and from interaction of all those quasiparticles. In our case we
take into account the\ double-degenerate vacuum by means of the shift:%
$$
x=\overline{x}-X, 
$$
then expand the integral around the two saddle points and apply the
self-similar renormalization to the resulting asymptotic expansion in powers
of a small parameter $g^{1/2},\ $continuing it to the region of $g\sim 1.$
Following these prescriptions, represent the integrand in the vicinity of
one of the saddle points in the form:%
$$
\exp (x^2-gx^4)\approx \exp (\frac 1{4g})\exp (-2X^2)\exp (A(g)X^3)\approx
\exp (\frac 1{4g})\exp (-2X^2)(1+A(g)X^3+...), 
$$
and perform the integration, so that 
\begin{equation}
\label{54}I(g)\approx 2\exp (\frac 1{4g})(a+bg^{1/2}+...),\qquad a=\sqrt{\pi 
}2^{-3/2},\quad b=2^{-3/2}.
\end{equation}
Applying the self-similar renormalization, we readily obtain 
\begin{equation}
\label{55}I^{*}(g)=2a\exp (\frac 1{4g})\exp (\frac{bg^{1/2}}a).
\end{equation}

Despite the absence of dynamics in the zero-dimensional case, $I^{*}(g)\ $%
consists of two factors of different nature; one of them is non-analytic in
the coupling constant, resembling the well-known ''instanton'' term within
the framework of nontrivial $D-$dimensional field theories, the second is
analytic in $g^{1/2}\ $and resembles a contribution from the excitations
above the instanton-anti-instanton background.

The percentage error for the renormalized $I^{*}(1),\ $calculated with
respect to the exact $I(1)=2.762,\ $is $2.462\%,$ and considerable
improvement is achieved compared to the percentage error of the perturbative
expansion (54), equal to$-8.834\%$.

\section{Strong Coupling Regime}

\subsection{Zero-Dimensional Case.}

Let us apply to (43) the so-called ''strong-coupling'' expansion, in powers
of $1/g,\ $with a quartic term of the integrand taken as a zero
approximation, representing the integrand of (43) as follows%
$$
\exp (-x^2-gx^4)\approx \exp (-gx^4)(1-x^2+\frac{x^4}2+...). 
$$
After integration we obtain the expansion in inverse powers of the coupling
constant 
\begin{equation}
\label{56}J(g)\approx ag^{-1/4}+bg^{-3/4}+cg^{-5/4}+...,\quad a=\frac{1.813}{
\sqrt{\pi }},\ b=\frac{-0.612}{\sqrt{\pi }},\ c=\frac{0.227}{\sqrt{\pi }}.
\end{equation}
We\ write down the following consecutive approximations to the quantity\ $%
J(z),\ $where $z=g^{-1/4},$%
$$
J_1(z)=az, 
$$
$$
J_2(z)=az+bz^3, 
$$
$$
J_3(z)=az+bz^3+cz^5\ ; 
$$
and the multiplier $m_2(z,s)=1+\frac ba\frac{3+s}{1+s}z^2\ $reaches its
minimal zero value at $s=0.019$, being much smaller than the minimal value
of the corresponding multiplier $m_3(z,s)=m_2(z,s)+\frac ca\frac{5+s}{1+s}%
z^4.\ $Therefore, the renormalized quantity $J_2^{*}(z),\ $will correspond
to the more stable trajectory, than $J_3^{*}(z).$ Following the standard
prescriptions of Sec.3 we obtain%
$$
J_2^{*}(x)=ax\{1-\frac{2b}{a[1+s(x)]}x^2\}^{-\{s(x)+1\}/2},\qquad s(x)=-
\frac{a+3bx^2}{a+bx^2}. 
$$
The percentage error for the renormalized quantity $J_2^{*}(1),\ $calculated
with respect to the exact $J(1)=0.772,\ $is $2.266\%,$ and a considerable
improvement is reached compared to the percentage error of the perturbative
expansion (53) with only starting two terms taken into account, equal to$%
-12.208\%$.

Represent (43) in a little bit different form: 
\begin{equation}
\label{57}J(g)=g^{-1/4}(a+bg^{-1/2}+cg^{-1}+...)\equiv g^{-1/4}[a+\overline{J%
}(g)],
\end{equation}
and write down the following set of approximations to $\overline{J}(g)$ ,
using the variable $y=g^{-1/2}:$%
$$
\overline{J_1}(y)=by, 
$$
$$
\overline{J_2}(y)=by+cy^2. 
$$
The multiplier $m_2(y,s)=1+\frac cb\frac{2+s}{1+s}y,$ is minimal at $s=0$,
and $\mid m_2(y,0)\mid <1.$ The renormalized quantity $\overline{J_2}^{*}(y)$
can be readily written down:%
$$
\overline{J_2}^{*}\ =\frac{by}{1-\frac cby} 
$$
Recalculating now $J^{*}(g),\ $we obtain $J^{*}(1)=0.771,$with the
percentage error $-0.13\%$, much better than the percentage error $4.386\%,\ 
$given by the initial expansion (57). Even at small $g=0.21$, the percentage
error given by the renormalized expression, remains less than $1\%$; at the
same time, the percentage error given by the initial expression reaches $%
43.538\%.$

\subsection{One-Dimensional Case.}

Consider the dimensionless ground state energy $e(g)$ of the celebrated
quantum one-dimensional quartic anharmonic oscillator, closely connected to
the so-called $\varphi ^4$ model in the quantum field theory (see e.g. [3]).
Here $g$ stands for the dimensionless coupling constant, expressed through
the parameters entering the Hamiltonian of the system by the known relation
(see e.g [3,41]). The asymptotic expansion for $e(g)$ in the strong coupling
limit, corresponding to $g\rightarrow \infty $, is known (see e.g.[41]) in
the following form: 
\begin{equation}
\label{58}e(g)\cong ag^{1/3}+bg^{-1/3}+cg^{-1},\quad a=0.667986,\
b=0.14367,\ c=-0.0088. 
\end{equation}
Let us, using the experience gained while considering the strong coupling
limit of the zero-dimensional field theory, renormalize the last two terms
of the expansion (58). Using the notation $y=g^{-1/3},$ we write down the
following set of approximations for the quantity $\overline{e}=e-ag^{1/3}:$%
$$
\overline{e}_1(y)=by, 
$$
$$
\overline{e}_2(y)=by+cy^3, 
$$
with the multiplier $m_2(y,s)=1+\frac cb\frac{3+s}{1+s}y^2,$ possessing
minimal value at $s=0$, when $g\succeq 0.1.$ The renormalized expression can
be obtained following the standard prescriptions of Sec.III. Returning to the
initial variable we obtain 
\begin{equation}
\label{59}e^{*}(g)=ag^{1/3}+\frac{b^{3/2}}{\sqrt{bg^{2/3}-2c}}. 
\end{equation}
An accuracy given by $e^{*}(g),$ can be elucidated by comparison with the
''exact'' numerical results (see e.g. [41]). At $g=0.3,$\ the percentage
error, given by (59), is equal to $-0.099\%$, at $g=1$ it is $-0.022\%$, and
at $g\rightarrow 200$ the percentage error tends to zero. To our knowledge, 
these results are better than those obtained by other analytical methods. 
On the other hand, at small $g=0.001,$ an accuracy of the formula (59) is 
by far inferior, compared to many other analytical methods. The reason can be
understood if notice, that $e^{*}(0)=0.41048,$ strongly deviating from the
known value $1/2$, but being much better than the infinite value predicted
by the initial expansion (58). We conclude, remarking, that using the
effective time $t^{*}\ $as an optimization parameter, determined from the
condition $e^{*}(0)=1/2,$ one can achieve better agreement of the
renormalized formulae with the exact results in the region of small coupling
constants. For the goal being pursued in the present paper, it is fairly
enough to limit the discussion by formulae (59), designed for the
intermediate and strong-coupling regimes.

\section{Equation of State}

\subsection{System of Hard Spheres.}

We demonstrate below, how the self-similar bootstrap can be applied in the
theory of equations of state for simple liquids. Consider the model 
system of hard spheres with the diameter $d$ [42,43], where the
empirical equation of state, connecting pressure $p,$ temperature $T,$ the
number density $n$ and the reduced density $\rho =\pi nd^2/6,$ is known: 
\begin{equation}
\label{60}\frac p{nkT}=\frac{1+\rho +\rho ^2-\rho ^3}{(1-\rho )^3}\ .
\end{equation}
The equation of state (60) agrees very well with the molecular dynamics
results [42]. On the other hand, the theoretical virial formula according to
Percus-Yevick [41,42], is given as follows: 
\begin{equation}
\label{61}\frac p{nkT}=\frac{1+\rho +\rho ^2-3\rho ^3}{(1-\rho )^3}\ .
\end{equation}
These two expressions almost coincide at low densities, e.g at $\rho =0.1,\ $%
the percentage error of Eq.(61), as compared with (60), equals $-0.18\%$ ;
while for the intermediate and high densities the agreement becomes very
poor, e.g. at $\rho =0.5$, the percentage error is $-15.385\%$, and at $\rho
=0.8\ \ $it equals $-53.112\%$ .

Consider the regular part of (61), defined as $r:$%
\begin{equation}
\label{62}r=1+\rho +\rho ^2-3\rho ^3, 
\end{equation}
as an asymptotic, low-density expansion for the ''true'' regular part $
\widetilde{r}(\rho )$,$\ $and try to continue the expression (62) from the
region of $\rho \ll 1,$ to the region of $\rho \preceq 1$.$\ $It seems
reasonable to use for renormalization only the last three terms from (62),
since the constant term describes the ideal gas behavior and we are
interested in the region of high densities.$\ $Let us\ write down the
following consecutive approximations to the quantity\ $\overline{r}=r-1:$%
$$
\overline{r_1}=\rho , 
$$
$$
\overline{r_2}=\rho +\rho ^2, 
$$
$$
\overline{r_3}=\rho +\rho ^2-3\rho ^3 
$$
The multipliers are%
$$
m_2(\rho ,s)=1+\rho \frac{2+s}{1+s}, 
$$
$$
m_3(\rho ,s)=m_2(\rho ,s)-3\rho ^2\frac{3+s}{1+s}, 
$$
$$
m_1^{\prime }(\rho ,s)=1-3\rho \frac{3+s}{2+s}. 
$$
It is admissible to apply here the self-similar bootstrap renormalization in
the form of the continued exponentials, since at every step of the procedure
the exponential summation is performed along the stable trajectory. $\ $%
Following the standard prescriptions of Sec.4, we obtain 
\begin{equation}
\label{63}\widetilde{r}(\rho )=\rho \exp (\rho \exp (-3\rho )). 
\end{equation}
The multiplier $M(\rho ),\ $corresponding to (63)$,$ is given by the
expression%
$$
M(\rho )=\exp [\rho \exp (-3\rho )]\exp (-4\rho )(1-3\rho ), 
$$
and is very small at $\rho >1/3,\ $e.g. $M(0.8)=-0.061,$ signaling the
robust stability of the sequence of the continued exponentials (63).
Recalculating 
\begin{equation}
\label{64}\widetilde{\frac p{nkT}}=\frac{\widetilde{r}(\rho )+1}{(1-\rho )^3}%
, 
\end{equation}
and comparing it to the empirical formula (60), we obtain that at $\rho
=0.1,\ $the percentage error equals $-0.118\%$ ; at $\rho =0.5,$ the
percentage error is $-4.061\%$, and at $\rho =0.8\ $it equals $-3.516\%.$

We see that the equation of state (64), obtained from the bootstrap
self-similar renormalization, much better and more uniformly agrees with the
computer experiment, than the initial virial expansion (61), over the entire
range of densities.The agreement drastically, by $17$ times, improves in the
region of high densities.

\subsection{ System of Hard Hexagons}

The model of ''hard hexagons'' represents a simple two-dimensional model of
impenetrable molecules on the triangular lattice. The model allows an exact
solution [44], and the phase transition from the liquid phase existing above
the critical value of the so-called activity $z_c=11.0917...,\ $to the solid
phase, existing below $z_c,$ is well studied. The equation of state,
describing the dependence of the order parameter $R$ \ on the
activity-related parameter, can be written down in quite a complicated and
not very convenient form [44]. On the other hand, the critical value of
density $\rho _c\ $at the point of phase transition is known too, and equals
$0.27639...$[44]. Independently, the high-density expansions of the order
parameter in powers of the inverse activity $z^{\prime }\equiv 1/z,$ or in
powers of the high-density variable $\rho ^{\prime }=1-3\rho ,$ were
obtained [45]. Their quality is considered as very high, since the critical
parameters could be determined from them with extremely high accuracy, using
the Pade approximants in conjunction with some extrapolation methods [45].
We present below simple expressions for the equation of state, obtained as a
continuation of the high-density expansions of the order parameter up to the
point of phase transition.

The expansion of the order parameter in powers of $\rho ^{\prime }$ is given
as follows [45]: 
\begin{equation}
\label{65}R=1-3(\rho ^{\prime })^2-9(\rho ^{\prime })^3-36(\rho ^{\prime
})^4-159(\rho ^{\prime })^5. 
\end{equation}

Let us apply to (65) the bootstrap self-similar renormalization based on the
exponential summation at every step, and leading to the equation of state
for the system of hard hexagons in the form of the continued exponentials: 
\begin{equation}
\label{66}\widetilde{R}(\rho ^{\prime })=\exp (-\rho ^{\prime }\exp (3\rho
^{\prime }\exp (3\rho ^{\prime }\exp (4\rho ^{\prime }\exp (\frac{53}{12}%
\rho ^{\prime }))))). 
\end{equation}
The function $\widetilde{R}(\rho ^{\prime })$ approaches zero at $\rho
_c^{\prime *}=0.170005(\pm 1),$ corresponding to the $\rho _c^{*}=0.276665,$
deviating from the exact value by $0.1\%$. Thus, the renormalized equation
of state (66), agrees with the initial expansion in the region of $\rho
^{\prime }\ll 1$ by design, and predicts the point of phase transition with
very high accuracy. On the other hand, the form of continued exponential,
can be justified {\it aposteriori, }analyzing the multipliers (42), where it
can be shown, after some lengthy, but routine calculations, that $M_5(\rho
^{\prime })\rightarrow 0,$ in the region of $\rho ^{\prime }\approx \rho
_c^{\prime *},\ $i.e. the stability condition is satisfied along the
trajectory, described by the sequence of approximations corresponding to
(66), in the vicinity of the critical point. Similarly, using the known
expansion of $R$ up to the fifth order terms in $z^{\prime },\ $the
corresponding equation of state can be obtained. In this case we found the
critical $z_c^{\prime *}\simeq 12.1803(\pm 1),$ deviating from the exact
value by $9.8\%.$

\subsection{Polymer Coil}

The expansion factor $\alpha $ of the polymer chain, within the framework of
a standard\ ''beads-on-string'' model, is conveniently represented as a
function $\alpha ^2=$$\alpha ^2(z)\ $of the parameter $z=2(\frac 3{2\pi
})^{3/2}N^{1/2}B/a^3,$ where $N$ \ is the total number of links in the
chain, $a$ stands for the typical distance between the beads-monomers, and $%
B $ is the second virial coefficient [46,47]. We consider below only the
case of a polymer coil, corresponding to $z>0.$ In the region of $z\ll 1,\ $%
the perturbation theory in powers of $z$ can be developed and for the
short-range potentials one can find [46,47] that 
\begin{equation}
\label{67}\alpha ^2=\alpha ^2(z)=1+k_1z+k_2z^2+...,\quad k_1\approx 1.28,\ \
k_2=-20.8. 
\end{equation}

One of the important problems in the physics of polymer coils, consists in
the continuation of the expansion (67) to the region of $z\sim 1\ $[46,47].
On the other hand, in the limit of $z\gg 1,$ $\alpha $ is related to $z$ by
a simple power-law 
\begin{equation}
\label{68}\alpha \sim z^{2\nu -1}, 
\end{equation}
where the critical index $\nu \geq 1/2,\ $can be calculated by different
methods\ [46-48]. We propose below, using the self-similar renormalization,
a simple way to continue (67) to the region of arbitrary $z,$ including both
known limiting cases and allowing to estimate $\nu \ $from the stability
condition. The problem of this type was already mentioned above, in Sec.II.
From the viewpoint of the applicability of the stability conditions, it is
worthwhile to study $\alpha ^{-2}(z)\equiv a(z),$ re-expanding it in powers
of $z,\ $so that 
\begin{equation}
\label{69}a(z)\approx 1+b_1z+b_2z^2+...,\quad b_1=-1.28,\ b_2=22.438. 
\end{equation}

The set of approximations to $a(z),\ $including the two starting terms from
(69), can be written down as follows:%
$$
a_0=1, 
$$
$$
a_1=1+b_1z, 
$$
and the expression for the renormalized quantity $a_1^{*}\ $can be readily
obtained: 
\begin{equation}
\label{70}a_1^{*}=(\frac{s_1}{s_1-b_1z})^{s_1}\Longrightarrow (\frac{s_1}{%
-b_1})^{s_1}\ z^{-s_1}\ (z\rightarrow \infty ),
\end{equation}
where the stabilizer $s_1$ should be positive, if we want to reproduce in
the limit of $z\rightarrow \infty ,$ the correct power-low behavior of the $%
\alpha ^2(z)$. A different set of approximations, not including into the
renormalization procedure the constant term from (69), has the form:%
$$
\overline{a_1}=b_1z, 
$$
$$
\overline{a_2}=b_1z+b_2z^2, 
$$
and applying the standard procedure, we obtain 
\begin{equation}
\label{71}a_2^{*}=1+b_1z[1-\frac{b_2z}{b_1(1+s_2)}]^{-(1+s_2)}%
\Longrightarrow (\frac{-b_2}{1+s_2})^{-(1+s_2)}\ b_1^{2+s_2}z^{-s_2}\
(z\rightarrow \infty ).
\end{equation}
Demanding now, that both (70) and (71) have the same power-law behavior at $%
z\rightarrow \infty ,$ we find that 
$$
s_2=s_1=2(2\nu -1). 
$$
Requiring now the fulfillment of the stability criteria for the two
approximations (70) and (71) in the form of the minimal-difference condition
(see Introduction), we obtain the condition on the {\it positive} stabilizer 
$s_1,\ $i.e$,$ $s_1$ should be determined from the {\it minimum }of the
expression $A:$ 
\begin{equation}
\label{72}A=\left| \{(\frac{-b_2}{1+s_1})^{-(1+s_1)}\ b_1^{(2+s_1)}-(\frac{%
s_1}{-b_1})^{s_1}\}\right| .
\end{equation}
The minimum of (72) does exist and is located at the point $s_1\approx 0.5,$
Correspondingly, the index $\nu $ is equal to $0.625$, in a reasonable
agreement with all other theoretical and experimental estimates of this
index [46-48].

As it was pointed out in Sec.2, the results may also be obtained, if the
self-similar renormalization is applied to the initial expansion (67), for
the sought function $\alpha ^2(z),$ although it is formally divergent at $%
z\rightarrow \infty $ and the stabilizer should become negative to describe
this divergence correctly. By simple substitution of the coefficients and
changing the criteria on minimum of (72) to the {\it maximum} of the
analogous expression $K:$%
\begin{equation}
\label{73}K=\left| \{(\frac{-k_2}{1+s})^{-(1+s)}\ k_1^{(2+s)}-(\frac
s{-k_1})^s\}\right| , 
\end{equation}
with respect to the now {\it negative \ }stabilizer $s\equiv 2(1-2\nu ).\ $%
One can see that the maximum is located at the point $s=-0.3719,$ leading to
the very reasonable estimate for the critical index $\nu =0.593.$$\ $Final
formulae have the following form: 
\begin{equation}
\label{74}\alpha _1^2(z)^{*}=(\frac s{s-k_1z})^s, 
\end{equation}
\begin{equation}
\label{75}\alpha _2^2(z)^{*}=1+k_1z[1-\frac{k_2z}{k_1(1+s)}]^{-(1+s)}\ . 
\end{equation}

Both formulae, (74) and (75), with $s\approx -0.3719$, can be used as an
approximate ''equation of state'' for the polymer in the whole range of the
parameter $z,$ satisfying, by design, both known virial and scaling
asymptotic expressions. \ 

\section{Critical Temperature of the 2D Ising Model from the Expansion
around the Dimension One}

In this Section we calculate the critical temperature $T_c\ $of the
two-dimensional ($2D)$ Ising model starting from the approximate expression
obtained within the framework of the so-called quasi-chemical approximation
[42]. This approximation gives $T_c$ as a function of the coordination
number $z:$%
\begin{equation}
\label{76}T_c(z)=\frac{-2}{\ln (1-2/z)}. 
\end{equation}
The expression (76) correctly describes the limit of the one-dimensional $%
(1D)\ $Ising model, with $T_c=0,\ $and at the infinite-dimensionality the
limit coincides with the well known Bragg-Williams result \ $T_c=z$. The
expansion around the latter limit has been widely used, although its
accuracy is not too good [49]. We adopt the different approach, expanding
(76), in powers of the parameter $z-2\equiv \Delta ,$ around its correct, $%
D=1\ (z=2),$ limit, and use the self-similar renormalization to continue the
expansion valid at $\Delta \ll 1\ $to the region of $\Delta =2,\ $%
corresponding to the $2D$ Ising model with the quadratic lattice. The
expansion of the inverse expression (76), up to the quadratic term in $%
\Delta ,\ $has the following form: 
\begin{equation}
\label{77}T_c^{-1}(\Delta )\approx \frac{\ln 2}2[1+\frac 1{\ln 2}\ln (\frac
1\Delta )]+\frac 14\Delta -\frac 1{16}\Delta ^2. 
\end{equation}
We renormalize separately the logarithmic 
\begin{equation}
\label{78}L(\Delta )=[1+\frac 1{\ln 2}\ln (\frac 1\Delta )] 
\end{equation}
and power-law contributions 
\begin{equation}
\label{79}P(\Delta )=\frac 14\Delta -\frac 1{16}\Delta ^2, 
\end{equation}
separating in this way, the effects of long-range and short-range
contributions to $T_c$.

The standard prescriptions of Sec.III are fully applicable to the expression
(78) containing the logarithmic term. Two consecutive approximations to $%
L(\Delta )$ are 
$$
L_0(\Delta )=1, 
$$
$$
L_1(\Delta )=1+\frac 1{\ln 2}\ln (\frac 1\Delta ), 
$$
with the expansion function \ $f=\Delta ^s,\ $and the multiplier 
$$
m_1(\Delta ,s)=1-\frac 1{\ln 2}[\ln (\Delta )+\frac 1s], 
$$
equal, at $\Delta =2,\ $to zero as $s\rightarrow \infty .$ The velocity
function has the following form:%
$$
v(s,f)=-\frac f{\ln 2}\frac{\ln f}s. 
$$
Calculating the evolution integral and taking the limit of $s\rightarrow
\infty ,\ $we obtain 
$$
L^{*}(\Delta )=\Delta ^{-1/\ln 2} 
$$
The expression for $P(\Delta )^{*},\ $can be readily written down in the
case of summation along the stable trajectory, corresponding$\ $to $%
s\rightarrow \infty $: 
$$
P^{*}(\Delta )=\frac 14\Delta \exp (-\frac 14\Delta ). 
$$
Recalculating $T_c^{*},$ we obtain $T_c^{*}(\Delta =2)=2.321.\ \ $The
percentage error equals $2.292\%$ , when compared to the exact Onsager
result $T_c=2.269.\ \ $It should be remembered, that the quasi-chemical
approximation (76) works with the percentage error of $27.149\%,$ and that
the initial expansion (77) deviates from the exact result by $76.289\%$.
Also, one of the best known approximate theoretical schemes, known as the
Kikuchi method [42], gives the percentage error equal to $6.831\%$.

\section{Temporal Asymptotes of the Diffusion Equation with Random
Stationary Noise}

\subsection{Poisson spectrum}

Consider the diffusion of particles in a medium with randomly distributed
traps, whose local density $\alpha ({\bf r})\ $is described by the
nonnegative Poisson random field [50-52]. The local particle density $n({\bf %
r},t)\ $in the presence of traps is described by the following equation: 
\begin{equation}
\label{80}-\frac \partial {\partial t}n({\bf r},t)=-\nabla ^2n({\bf r}%
,t)+\alpha ({\bf r})n({\bf r},t), 
\end{equation}
where we put the diffusion coefficient equal to one.

This problem is formally equivalent to the Shroedinger equation with
imaginary time and potential $\alpha ({\bf r})\ $[50-52]$.\ $The eigenvalues
of the Shroedinger operator corresponding to (80), $E_i$, are nonnegative.
In addition, the density of states $p(E),\ $near the finite fluctuational
limit of the spectrum, located at $E=0,\ $and$\ $formed due to the rare
fluctuations of the potential with $\alpha $ close to zero in large regions
of space, is known explicitly: 
\begin{equation}
\label{81}p(E)=f(E)\exp (-E^{-D/2}), 
\end{equation}
where $D\ $is the dimensionality of space [53,54]. Hereafter, for the sake
of simplicity, we omit the constant terms in the exponential for the density
of states. Also, for $D=1,$ the pre-exponential factor $f(E)\sim E^{-3/2},$
is known [53] and, because of this, we consider below only the
one-dimensional case. The general solution of (80) can be readily written
down [52] in terms of the eigenfunctions and eigenvalues of the
corresponding quantum-mechanical problem: 
\begin{equation}
\label{82}n(r,t)=\sum_ic_i\psi _i(r)\exp (-E_it), 
\end{equation}
and, considering only contributions from the rare fluctuations of the
potential, the mean density $\langle n\rangle $ over the entire volume for $%
t\rightarrow \infty \ $can be represented in the form of the integral: 
\begin{equation}
\label{83}\langle n(t)\rangle =n_0\int_0^\infty p(E)\exp (-Et)dE, 
\end{equation}
where $n_0\ $describes the initial homogeneous particle distribution. This
integral can be evaluated by the method of steepest descent and the leading
exponential term $\sim \exp (-t^{1/3})$ can be written down [49-51]. Similar
estimates were obtained for arbitrary $D\ $[50-52].

We will obtain below the higher-order contributions to the asymptotic
expansion of (83) near the saddle point $\overline{E}=(\frac 1{2t})^{2/3}$
and, using the self-similar renormalization, obtain the leading corrections,
as $t\rightarrow \infty $, to the pre-exponential factor $f(\overline{E}%
)\sim t^{-1}.\ $

Let us represent $E\ $ in the vicinity of $\overline{E}$ as $E=$ $\overline{E%
}+\epsilon \ $and expand the expression 
$$
\Phi (t,\epsilon )\equiv \ln [p(E)\exp (-Et)] 
$$
in powers of $\epsilon $ up to the third order terms, so that%
$$
\Phi (t,\epsilon )\approx -at^{1/3}-A(t)\epsilon ^2+B(t)\epsilon ^3, 
$$
$$
a=3\times 2^{-2/3},\quad A(t)=3\times 2^{-4/3}t^{5/3},\quad B(t)=5\times
2^{-5/3}t^{7/3} 
$$
and, expand the$\ \exp \{B(t)\epsilon ^3\}$ in powers of $\epsilon ,\ $so
that%
$$
\exp \{\Phi (t,\epsilon )\}\approx \exp \{-at^{1/3}\}\exp \{-A(t)\epsilon
^2\}\times [1+B(t)\epsilon ^3+...]. 
$$
Now, $\langle n(t)\rangle $ can be written down as follows: 
\begin{equation}
\label{84}\langle n(t)\rangle \sim t^{1/6}\exp
\{-at^{1/3}\}[1+bt^{-1/6}],\qquad b=0.684, 
\end{equation}
i.e, the corrections to the pre-exponential factor are obtained in the form
of an expansion in inverse powers of $t$, valid as $t\rightarrow \infty .$
Our aim is to continue this expression to the region of $t\sim 1.$

Apply now the self-similar renormalization to the quantity $\overline{n}%
(t)=1+bt^{-1/6},\ $with the two consecutive approximations:%
$$
\overline{n}_0=1, 
$$
$$
\overline{n}_1=1+bt^{-1/6}. 
$$
Following the standard prescriptions, the renormalized quantity $\overline{n}%
^{*}(t)$ can be obtained: 
\begin{equation}
\label{85}\overline{n}^{*}(t)=\{t^{1/6}+\frac b{6s(t)}\}^{6s(t)}, 
\end{equation}
where the stabilizer 
$$
s(t)=\frac b{6t^{1/6}+6b}, 
$$
is defined as zero of the multiplier 
$$
m_1(s,t)=1+\frac{bt^{-1/6}(s-\frac 16)}s. 
$$
For the intermediate region $1\ll t<\infty ,$ the simple expression can be
written down:%
$$
\overline{n}^{*}(t)\propto t^{b/6t^{1/6}}, 
$$
which gives the correction to the pre-exponential factor in the form of
continued noninteger powers. It is worth noting, that already the starting
terms of the asymptotic expression (85), lead to the approximation-cascade
trajectory with zero multiplier.

\subsection{ Gaussian spectrum.}

Consider equation of the same type that (80), with the only difference that
the Poisson potential $\alpha $ is replaced by the random potential $U({\bf r%
})\ $with the properties of the Gaussian ''white noise'', i.e. $\overline{U(%
{\bf r})}=0,\ \langle U({\bf r})U({\bf r}^{^{\prime }})\rangle \propto
\delta ({\bf r-r}^{^{\prime }}):$%
\begin{equation}
\label{86}-\frac \partial {\partial t}n({\bf r},t)=-\nabla ^2n({\bf r},t)+U(%
{\bf r})n({\bf r},t).
\end{equation}
Equation of this type, but with a noise dependent both on space and time,
can be easily transformed to a nonlinear Burgers equation,
Kardar-Parisi-Zhang (KPZ) equation, and also it describes some other closely
related physical problems [55]. Stationary random potential is not
meaningless within the framework of, say, KPZ where one can think about the
stationary, random in space, perturbations of a growing interface. It is
also considered in biology as a model for population dynamics in the
presence of a random distribution of food [55]. Equation of the type (86)
with a random potential $U({\bf r})\ $can be also transformed to the
corresponding Shroedinger equation with imaginary time. Therefore, one can
use the known properties of the spectrum of the corresponding
quantum-mechanical problem to study the $t\rightarrow \infty $ behavior of
the diffusion equation with stationary, randomly distributed sources and
sinks. The fluctuational limit of the spectrum is situated now at $%
E\rightarrow -\infty ,$ with the exponentially small density of states in
its vicinity 
\begin{equation}
\label{87}p(E)=f(E)\exp (-\mid E\mid ^{2-D/2}),
\end{equation}
being formed due to the rare fluctuations of the potential with large
negative values, separated from each other by the distances much larger than
their own sizes [53]. Such fluctuations may be again considered separately.
The eigenvalues corresponding to the eigenfunctions localized at these
fluctuations are now negative, in distinction from the case considered above
and the mean density evolution as $t\rightarrow \infty \ $can be estimated
from the following integral: 
\begin{equation}
\label{88}\langle n(t)\rangle =n_0\int_0^\infty p(E)\exp (\mid E\mid t)dE.
\end{equation}

We again use the method of steepest descents and follow literally the same
steps as in the previous Subsection. The saddle point does exist for $\
1\leq D$$<4\ $(except at $D=2,\ D=4,$ where the situation becomes trivial)
and is given by the expression%
$$
\mid \overline{E\mid }=(\frac{2t}{4-D})^{2/(2-D)}\ . 
$$
The leading exponential term has the form 
$$
\langle n(t)\rangle \sim \exp \{a(D)t^{\frac{4-D}{2-D}}\},\qquad a(D)=(\frac{%
2-D}{4-D})(2-\frac D2)^{\frac 2{D-2}}, 
$$
with radically different behavior for $D=1\ $and $D=3:$%
\begin{equation}
\label{89}\langle n(t)\rangle \sim \exp \{\mid a(1)\mid t^3\},\qquad D=1, 
\end{equation}
\begin{equation}
\label{90}\langle n(t)\rangle \sim \exp \{-\mid a(3)\mid t^{-1}\},\qquad
D=3, 
\end{equation}
corresponding to an anomalously fast growth, compared to the simple $\exp
(t),$ and, anomalously slow decay, compared to $\exp (-t)$, respectively. We
think that this difference takes its origin from the principally different
properties of the corresponding Shroedinger operator, where it is known,
that at $D=1\ $all\ states of the particle are localized, while at $D=3$,
generally speaking, both localized and delocalized states are present [53].
These basic theorems, when applied to the case of diffusion, explain why at $%
1D$ , the random distribution of sources and sinks, causes an explosive
instability of the density fluctuations, while in $3D$, the disorder can
cause only longer decay times for the density fluctuations. Of course, the
instability can be cured by nonlinear interactions, that should be now
taken into account.

At $D=1,$ where $f(E)\sim E\ $[54]$,\ $applying the procedure already
discussed above, we obtain the expansion in the vicinity of the saddle point:%
$$
\langle n(t)\rangle \propto \exp \{\mid a(1)\mid t^3\}\times
t^{5/2}[1+bt^{-3/2}+...],\qquad b=\frac 1{2\sqrt{\pi }}, 
$$
and the renormalized expression can be readily written down, using the same
definition for $\overline{n}(t)$ as above: 
\begin{equation}
\label{91}\overline{n}^{*}(t)=\{t^{3/2}+\frac{3b}{2s(t)}\}^{\frac{2s(t)}%
3}\sim t^{3bt^{-3/2}/2},\qquad s(t)=\frac{3b}{2(b+t^{3/2})}. 
\end{equation}

At $D=3,\ $a little bit different situation occurs, since%
$$
\Phi (t,\epsilon )\approx -\mid a(3)\mid t^{-1}+A(t)\epsilon ^2-B(t)\epsilon
^3, 
$$
$$
\quad A(t)=t^3,\quad B(t)=2t^5, 
$$
and, in order to guarantee the convergence of the integrals, it is
reasonable to expand $exp(\Phi (t,\epsilon ))$ as follows: 
$$
\exp \{\Phi (t,\epsilon )\}\approx \exp \{-\mid a(3)\mid t^{-1}\}\exp
\{-B(t)\epsilon ^3\}\times [1+A(t)\epsilon ^2+...], 
$$
and for the $\langle n(t)\rangle $ we obtain%
$$
\langle n(t)\rangle \sim \exp \{-\mid a(3)\mid
t^{-1}\}t^{-5/3}[1+ct^{-4/3}+...],\qquad c\approx 0.296. 
$$
The renormalized expression for the pre-exponential factor has the following
form: 
\begin{equation}
\label{92}\overline{n}^{*}(t)=\{t^{4/3}+\frac{4b}{3s(t)}\}^{4s(t)/3}\sim
t^{4ct^{-4/3}/3},\qquad s(t)=\frac{4c}{3(c+t^{4/3})}. 
\end{equation}
We have demonstrated in this Section, that the self-similar renormalization
can be applied to the dynamical problems as well, generating the expressions
for the pre-exponential factors in the form of continued noninteger powers.

\section{Conclusion}

We suggested here a new variant of the self-similar approximation theory
permitting an easy and accurate summation of divergent series. The method is
based on a power-law algebraic transformation leading to an effective
increase of the order of perturbative terms. The powers of this
transformation play the role of control functions governing the convergence
of renormalized series. These control functions are defined by the principle
of maximal stability, i.e., from the minimization of mapping multipliers.
Such stabilizing control functions may be called stabilizers.

Another important novelty of the method is the multiple self-similar
renormalization converting all series into closed self-similar expressions.
This multiple and complete renormalization is called self-similar bootstrap.
The resulting effective sum of a divergent series can be presented through
analytical expressions containing exponentials and fractions, rational or
irrational. In particular cases, this can be only exponentials or only
fractions, depending on the behavior of control functions which dictate the
resulting form. Because of much larger variety of such resulting forms, the
method allows to present the answers in relatively simple analytical
expressions having at the same time quite high accuracy. The use of several
types of functions, such as exponentials and various fractions,
distinguishes this method from, e.g., the Pade$^{^{\prime }}\ $approximants
involving solely rational functions.

In order to prove that the suggested method really gives quite simple and
accurate expressions for the effective sums of divergent series, we , first
of all, considered several toy models cartooning the generating functionals
in field theory or partition functions in statistical physics. We
illustrated by these examples that the method works well in different
situations, for single-well and for double-well models, for weak and strong
coupling.

To stress the generality of the method, we applied it to several problems of
statistical physics of quite different nature: to constructing the equation
of state, to calculating the critical temperature, and to finding the time
asymptotics for stochastic dynamical processes. We hope that these various
and very different applications demonstrate well the validity of the method.

\vspace{0.5cm}

{\bf Acknowledgement}

\vspace{2mm}

Financial support of the National Science and Technology Development Council
of Brazil is appreciated.

\end{document}